\documentclass{elsart3}

\usepackage{amsmath}
\usepackage{graphicx}

\begin{document}

\begin{frontmatter}



\title{Quantum dynamics of exchange biased Single Molecule Magnets}


\author[label1]{W. Wernsdorfer}
\author[label2]{N. Aliaga-Alcalde}
\author[label1]{R. Tiron}
\author[label3]{D. N. Hendrickson}
\author[label2]{G. Christou}

\address[label1]{Lab. Louis N\'eel, CNRS, BP 166, 38042 Grenoble Cedex 9, France}
\address[label2]{Dept. of Chemistry, University of Florida, Gainesville, Florida 32611-7200}
\address[label3]{Dept. of Chemistry and Biochemistry, University of California at San Diego, La Jolla, California 92093-0358}

\begin{abstract}
We present a new family of exchange biased Single Molecule Magnets 
in which antiferromagnetic coupling between the two components 
results in quantum behaviour different from that of the individual 
SMMs. Our experimental observations and theoretical analysis suggest 
a means of tuning the quantum tunnelling of magnetization in SMMs. 
\end{abstract}

\end{frontmatter}

Various present and future specialized applications of magnets 
require monodisperse, small magnetic particles, and the discovery of 
molecules that can function as nanoscale magnets was an important 
development in this regard~\cite{Sessoli93b}. 
These molecules act as single-domain 
magnetic particles that, below their blocking temperature, exhibit 
magnetization hysteresis, a classical property of macroscopic 
magnets. Such single-molecule magnets (SMMs)~\cite{Aubin96} 
straddle the interface 
between classical and quantum mechanical behaviour because they also 
display quantum tunnelling of magnetization
~\cite{Novak95,Friedman96,Thomas96,Sangregorio97,Aubin98,Price99,Yoo_Jae00} 
and quantum phase interference~\cite{WW_Science99,Garg93}. 
Quantum tunnelling of magnetization can be advantageous 
for some potential applications of SMMs, for example, in providing 
the quantum superposition of states required for quantum computing. 
However, it is a disadvantage in other applications, such as 
information storage, where it would lead to information loss. Thus it 
is important to both understand and control the quantum properties of 
SMMs. Here we present a new family of 
supramolecular SMM dimers~\cite{WW_Nature02} 
in which antiferromagnetic coupling between the two components 
results in quantum behaviour different from that of the individual 
SMMs. Our experimental observations and theoretical analysis suggest 
a means of tuning the quantum tunnelling of magnetization in SMMs.

\begin{figure}
\begin{center}
\includegraphics[width=.4\textwidth]{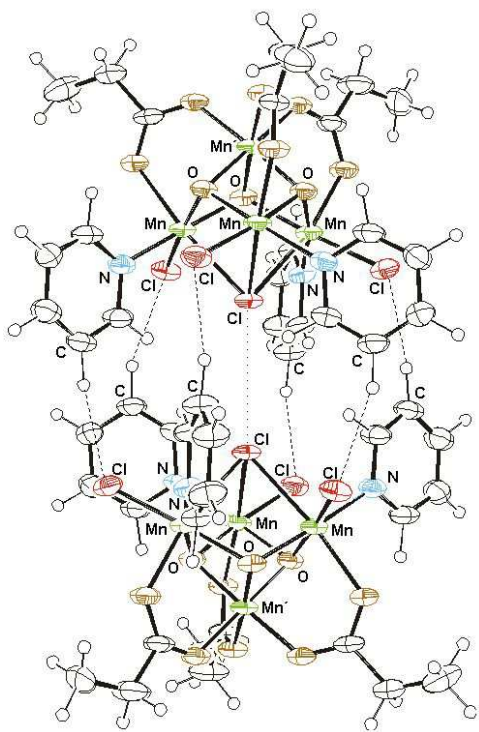}
\caption{
The structure of the [Mn$_4$]$_2$ dimer of 
[Mn$_4$O$_3$Cl$_4$(O$_2$CEt)$_3$(py)$_3$]. 
The small circles are hydrogen atoms. 
The dashed lines are C-H$\cdot\cdot\cdot$Cl hydrogen bonds and 
the dotted line is the close Cl$\cdot\cdot\cdot$Cl approach. 
The labels Mn and Mn' refer to Mn$^{\rm III}$ and Mn$^{\rm IV}$ ions, 
respectively.
}
\label{dimer}
\end{center}
\end{figure}

The compound has the formula

[Mn$_4$O$_3$Cl$_4$(O$_2$CEt)$_3$(py)$_3$] 
(hereafter Mn$_4$).  The preparation, 
X-ray structure, and detailed physical characterization 
have been reported~\cite{Hendrickson92,Nuria03}. 
The complex has a distorted cubane-like core geometry and 
is Mn$_3^{\rm III}$Mn$^{\rm IV}$. 
Strong superexchange coupling between the spins
leads to a well isolated $S = 9/2$ ground state.
Mn$_4$ crystallizes in the hexagonal space 
group R3(bar) with pairs of Mn$_4$ molecules 
lying Ôhead-to-headÕ on a crystallographic S6 
symmetry axis (Fig. 1). 
Thus, each M$_4$ has C$_3$ symmetry and the [Mn$_4$]$_2$ 
dimer has S$_6$ symmetry. 
This supramolecular arrangement is held together by six 
C-H$\cdot\cdot\cdot$Cl hydrogen bonds (dashed lines in Fig. 1) 
between the pyridine (py) rings on one Mn$_4$ 
and Cl ions on the other. The C-H$\cdot\cdot\cdot$Cl hydrogen bonds 
in [Mn$_4$]$_2$ have C$\cdot\cdot\cdot$Cl distances and C-H$\cdot\cdot\cdot$Cl angles 
of 0.371 nm and 161.57$^{\circ}$, respectively, typical of this type of 
hydrogen bond. 
The [Mn$_4$]$_2$ dimer also brings the central 
bridging Cl ions of each Mn$_4$ close together 
(0.3858 nm, dotted line in Fig. 1), almost at their 
Van der Waals separation (0.36 nm). Each [Mn$_4$]$_2$ is rather 
well separated from neighboring dimers.
The supramolecular linkage within [Mn$_4$]$_2$ introduces 
exchange interactions between the M$_4$ molecules 
via both the six C-H$\cdot\cdot\cdot$Cl pathways and 
the Cl$\cdot\cdot\cdot$Cl approach. 
All these seven interactions are expected to be weak, 
but combined they will lead to  noticeable antiferromagnetic
coupling between the Mn$_4$ units.

Antiferromagnetic coupling between the M$_4$ units 
in [Mn$_4$]$_2$ results in the latter being an excellent 
candidate for the study of quantum tunnelling 
in a system of truly identical, antiferromagnetically-coupled 
particles with a ground state S = 0. Such a system has long 
been sought for study after the theoretical conclusion~\cite{Barbara90} 
that tunnelling in antiferromagnets is much more pronounced 
than in ferromagnets. 

\begin{figure}
\begin{center}
\includegraphics[width=.4\textwidth]{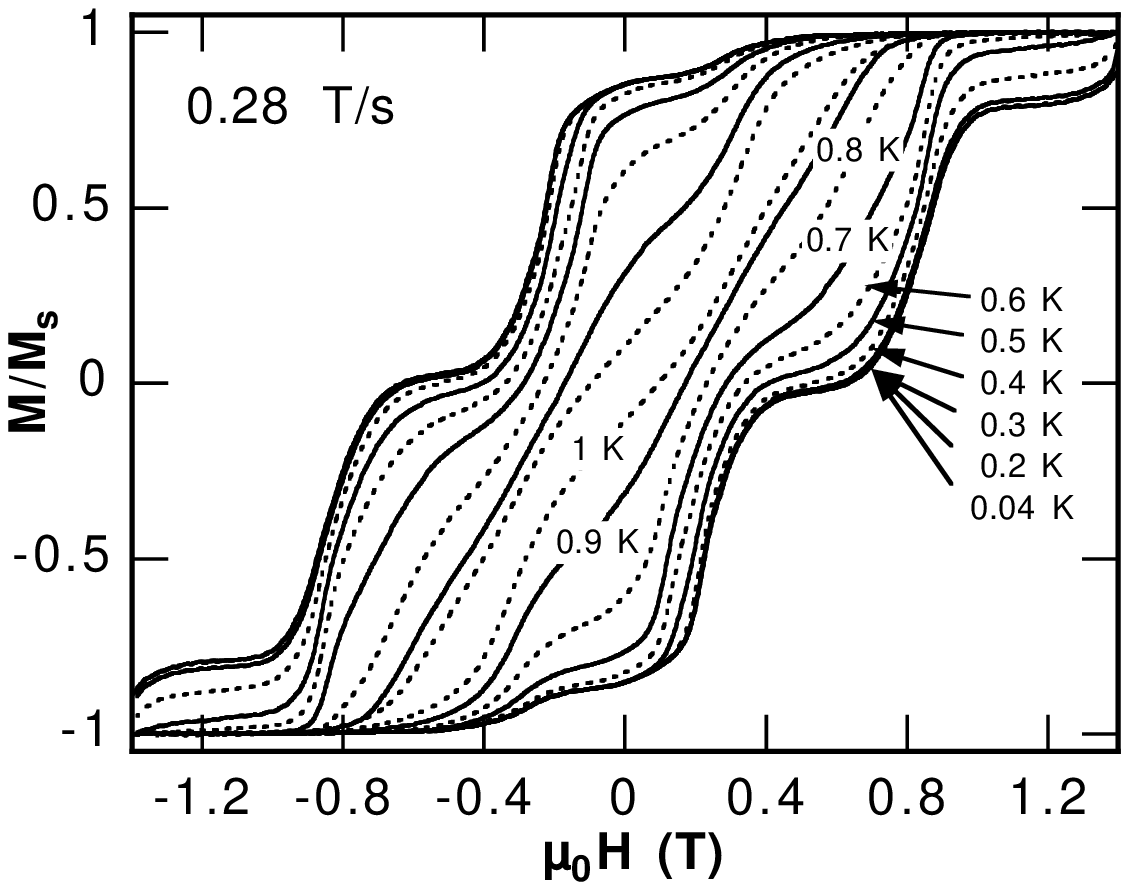}
\includegraphics[width=.4\textwidth]{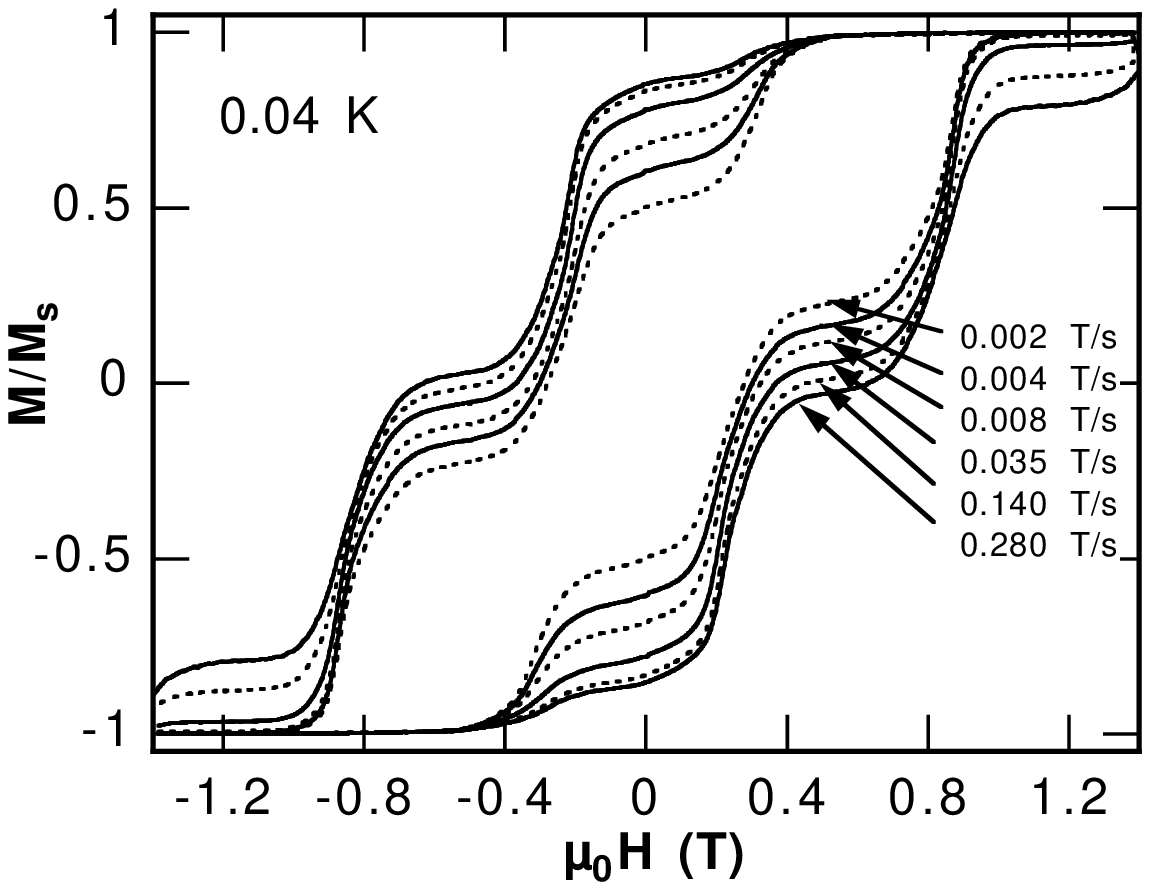}
\caption{
Normalized magnetization versus applied magnetic field. 
The resulting hysteresis loops are shown at (a) different temperatures 
and (b) different field sweep rates. 
Note that the loops become temperature-independent 
below about 0.3 K but are still sweep-rate-dependent 
owing to resonant quantum tunnelling between discrete 
energy states of the[Mn$_4$]$_2$ dimer.}
\label{hyst}
\end{center}
\end{figure}

Tunnelling studies on [Mn$_4$]$_2$ were performed 
by magnetization measurements on single crystals 
using an array of micro-SQUIDs~\cite{WW_ACP_01}. 
Fig. 2 shows typical hysteresis loops in magnetization versus
magnetic field scans with the field applied nearly along 
the easy axis of magnetization of [Mn$_4$]$_2$, that is, parallel to the 
S$_6$ axis~\cite{note1}.
These loops display step-like features separated by plateaus. 
The step heights are temperature-dependent above 0.3 K; 
below this, the loops become temperature-independent. 
As discussed below, the steps are due to resonant 
quantum tunnelling of the magnetization (QTM) between 
the energy states of the [Mn$_4$]$_2$ dimer. 

QTM has been previously observed for several 
SMMs~\cite{Novak95,Friedman96,Thomas96,Sangregorio97,Aubin98,Price99,Yoo_Jae00},
but the novelty for [Mn$_4$]$_2$ is that the QTM is now 
the collective behavior of the complete $S = 0$ dimer 
of exchange-coupled $S = 9/2$ Mn$_4$ quantum systems. 
This coupling is manifested as an exchange bias 
of all tunnelling transitions, and the resulting 
hysteresis loop consequently displays unique features, 
such as the absence for the first time in a SMM of a QTM 
step at zero field.

Before interpreting the hysteresis loops further, 
we discuss a simplified spin Hamiltonian 
describing the [Mn$_4$]$_2$ dimer. 
Each M$_4$ SMM can be modeled as a Ôgiant spinÕ of $S = 9/2$ 
with Ising-like anisotropy. The corresponding Hamiltonian is given by
\begin{equation}
	\mathcal{H}_i = -D S_{z,i}^2 + \mathcal{H}_{{\rm trans}, i} 
	+ g \mu_{\rm B} \mu_0 \vec{S_i}\cdot\vec{H} 
\label{eq_Hi}
\end{equation}
where $i = 1$ or 2 (referring to the two M$_4$ SMMs of the dimer), 
$D$ is the uniaxial anisotropy constant, 
and the other symbols have their usual meaning. 
Tunnelling is allowed in these half-integer ($S = 9/2$) 
spin systems because of a small transverse anisotropy $\mathcal{H}_{{\rm trans}, i}$ 
contains $S_{x,i}$ and $S_{y,i}$  spin operators and transverse 
fields ($H_{x}$ and $H_{y}$). 
The exact form of $\mathcal{H}_{{\rm trans}, i}$  is not important 
in this discussion. The last term in Eq. 1 is the 
Zeeman energy associated with an applied field. 
The Mn$_4$ units within the [Mn$_4$]$_2$ dimer 
are coupled by a weak superexchange interaction $J$ via 
both the six C-H$\cdot\cdot\cdot$Cl pathways and the Cl$\cdot\cdot\cdot$Cl approach. 
Thus, the Hamiltonian ($\mathcal{H}$) for [Mn$_4$]$_2$ is
\begin{equation}
	\mathcal{H} = \mathcal{H}_1 + \mathcal{H}_2 + J 
	\vec{S_1}\cdot\vec{S_2} 
\label{eq_H}
\end{equation}
where $S_1 = S_2 = 9/2$. 
The $(2S_1+1)(2S_2+1) = 100$ energy states of the dimer 
can be calculated by exact diagonalization and are plotted 
in Fig. 3a as a function of applied field along the easy axis. 
Each state of [Mn$_4$]$_2$ can be labelled by two quantum 
numbers $(M_1,M_2)$ for the two Mn$_4$ SMMs, 
with $M_1 = -9/2, -7/2, ..., 9/2$ and $M_2 = -9/2, -7/2, ..., 9/2$. 
For the sake of simplicity, we do not discuss the mixing
with other spin states that is due to transverse anisotropy terms,
that is $\mathcal{H}_{{\rm trans}, i} = 0$.

\begin{figure}
\begin{center}
\includegraphics[width=.4\textwidth]{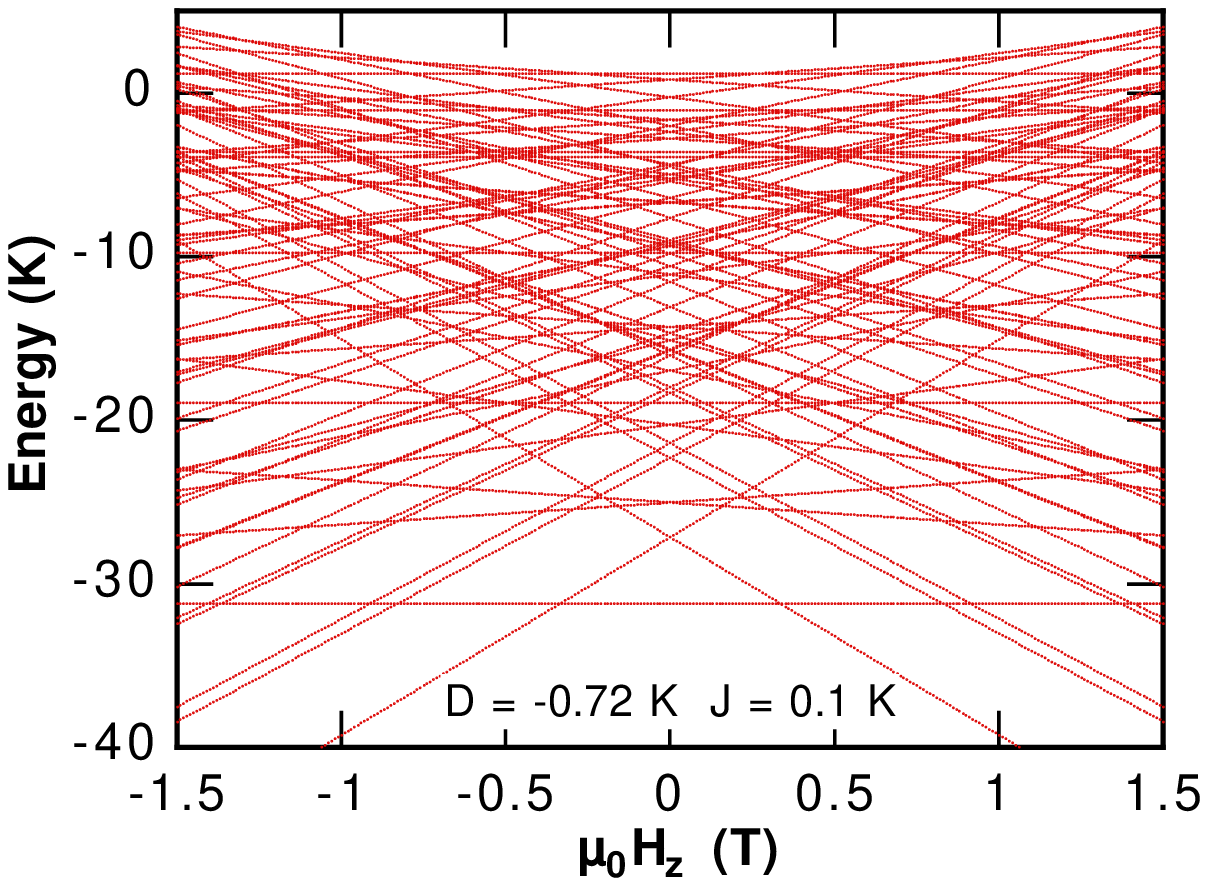}
\includegraphics[width=.4\textwidth]{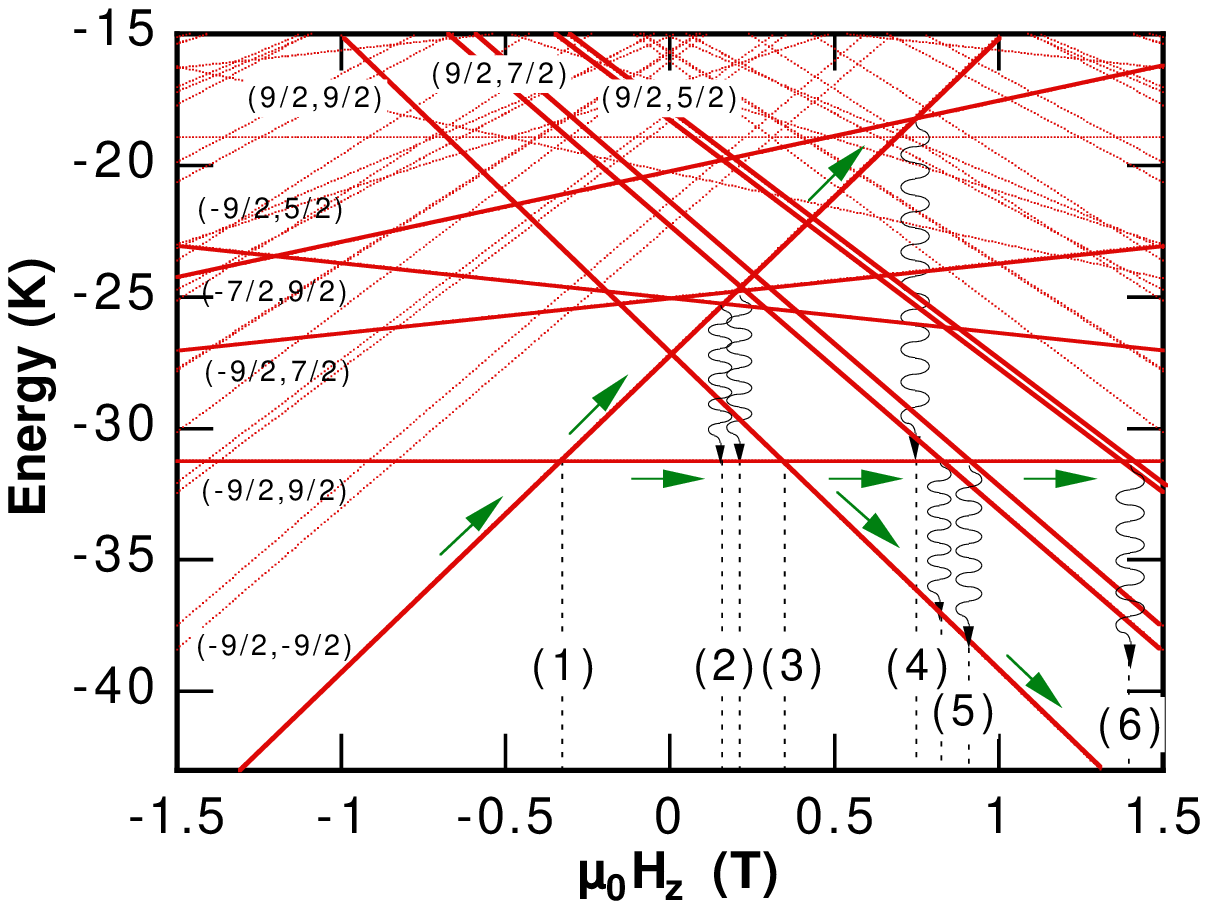}
\caption{
The spin state energies of [Mn$_4$]$_2$ as a function of 
applied magnetic field. 
(a) Energy vs field plot for the 100 states of 
the exchange-coupled dimer of two spin $S = 9/2$ Mn$_4$ units. 
Transverse terms in $\mathcal{H}_{{\rm trans}, i}$ are 
neglected for the sake of simplicity. However, transverse
exchange coupling terms are included.
(b) Enlargement of Fig. 3a showing only levels 
populated at very low temperature. 
Dotted lines, labelled (1) to (6), indicate 
the strongest tunnel resonances: 
(1) (-9/2,-9/2) to (-9/2,9/2); 
(2) (-9/2,-9/2) to (-9/2,7/2), followed by relaxation to (-9/2,9/2); 
(3) (-9/2,9/2) to (9/2,9/2); 
(4) (-9/2,-9/2) to (-9/2,5/2), followed by relaxation to (-9/2,9/2); 
(5) (-9/2,9/2) to (7/2,9/2), followed by relaxation to (9/2,9/2). 
(6) (-9/2,9/2) to (5/2,9/2), followed by relaxation to (9/2,9/2). 
For clarity, degenerate states such as (M,M') and (M',M) are not both listed.}
\label{Zeeman}
\end{center}
\end{figure}

At very low temperature, most of the excited spin 
states are not populated and can be neglected. 
Thus, Fig. 3b shows only the low-lying states involved 
in the magnetization reversal at very low temperature 
when sweeping the field from a high negative field 
to a positive one. At high negative field, the initial 
state is $(M_1,M_2) = (-9/2,-9/2)$, that is, both molecules 
are in the negative ground state. As the field is 
swept, the first (avoided) level crossing is at -0.33 Tesla. 
There is a non-zero probability, which depends upon 
the sweep rate, for the magnetization to tunnel from 
$(-9/2, -9/2)$ to $(-9/2, 9/2)$. 
(For convenience, we do not list here (or below) both 
degenerate $(M,M')$ and $(M',M)$ states). The smaller 
the sweep rate the larger is the tunnelling probability, 
i.e. the larger is the resulting step in Fig. 2b. 
At field values away from the avoided level crossing, 
the dimer states are frozen by the significant 
magnetic anisotropy barrier.

At the next level crossing at 0 T, there is the 
possibility of tunnelling from $(-9/2, -9/2)$ to $(9/2, 9/2)$, 
but this requires both Mn$_4$ molecules of the dimer 
to tunnel simultaneously. The corresponding tunnelling 
probability is very small and can hardly be seen in Fig. 2. 
At the next level crossing at 0.2 T, the dimer can 
undergo tunnelling from $(-9/2,-9/2)$ to $(-9/2,7/2)$, 
corresponding to one of the Mn$_4$ molecules tunnelling 
from the ground state to an excited state, 
followed by rapid relaxation from $(-9/2, 7/2)$ to 
$(-9/2, 9/2)$ (curly arrow in Fig. 3b). At the next 
level crossing at 0.33 T, the situation is analogous 
to that at -0.33 T, and the dimer can undergo 
tunnelling from $(-9/2, 9/2)$ to $(9/2,9/2)$.
Between 0.33 and 0.75 T, there are several level 
crossings from $(-9/2,-9/2)$ to excited states 
that require simultaneous 
tunnelling in both Mn$_4$ units and can be neglected. 
The level crossing at 0.75 T allows tunnelling from 
$(-9/2, -9/2)$ to $(-9/2, 5/2)$, followed by relaxation 
from $(-9/2, 5/2)$ to $(-9/2, 7/2)$ and $(-9/2, 9/2)$. 
At 0.87 T tunnelling can occur from $(-9/2, 9/2)$ to 
$(7/2, 9/2)$, followed by relaxation from $(7/2, 9/2)$ to $(9/2, 9/2)$.
Finally, at 1.4 T tunnelling can occur from $(-9/2, 9/2)$ to 
$(5/2, 9/2)$, followed by relaxation from $(5/2, 9/2)$ to $(9/2, 9/2)$.
The values of $D$ and $J$ used in Fig. 3 were calculated 
from the field positions of the steps in the 
hysteresis loops, ignoring transverse (and fourth order) 
anisotropy terms. The calculated values are $D \approx$ -0.72 K 
and $J \approx$ +0.1 K. The former is very close to those 
determined experimentally for several (isolated) 
Mn$_4$ SMMs~\cite{Aubin98b,Andres00}, and the latter is antiferromagnetic 
(positive value) and very weak, as expected for an exchange 
interaction via the six C-H$\cdot\cdot\cdot$Cl and the Cl$\cdot\cdot\cdot$Cl pathways. 
A value of $J \approx 0.1$ K was also obtained from 
susceptibility determinations using hysteresis loops 
measured in the 1 to 8 K range (Fig. 4).

\begin{figure}
\begin{center}
\includegraphics[width=.4\textwidth]{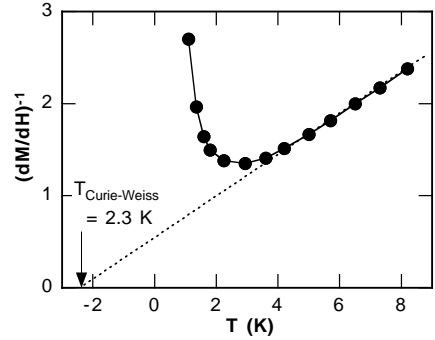}
\caption{
Inverse zero field susceptibility, $dM/dH^{-1}(H = 0)$, 
versus temperature, determinated using hysteresis loops.
The negative Curie-Weiss constant of 2.3 K is in good
agreement with an antiferromagnetic coupling of $J \approx 0.1$ K.}
\label{chi}
\end{center}
\end{figure}

We tried to get a deeper insight into the relaxation 
processes by relaxation measurements at constant 
applied field for initially saturated or
demagnetized magnetization (Fig. 5). The
relaxation paths of the former are analogous to
those of hysteresis loops (Fig. 2) whereas
the latter starts in an initial state of $(-9/2, 9/2)$
for nearly all dimers. Thus, transitions (2) 
and (4) should
not occur that can be seen well in Fig. 5. 
Furthermore, this shows that only few dimers
relaxe via transitions (4) of initially saturated
magnetization.
Note that all transitions 
are not well resolved, due to significant broadening 
by dipolar effects and small exchange coupling between
neighboring dimers.

\begin{figure}
\begin{center}
\includegraphics[width=.4\textwidth]{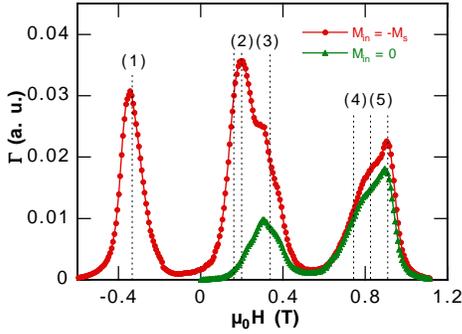}
\caption{
Relaxation rate versus applied field for initially saturated or
demagnetized magnetization. The dominating tunnel transitions 
are indicated by dashed lines and numbers (1) to (5), 
which are indicated in Fig.3b and explained in the text. 
The dashed lines are at positions calculated from Fig. 3b.}
\label{gamma}
\end{center}
\end{figure}

Relaxation data were determined from DC relaxation 
decay measurements: first a field of 1.4 T was applied 
to the sample at 5 K to saturate the magnetization 
in one direction, and the temperature lowered to a 
chosen value between 1 and 0.04 K. The field was 
then swept to 0 T at 0.28 T/s and the magnetization 
in zero field measured as a function of time (Fig. 6). 
From these data, relaxation times could be 
extracted which allows the construction of 
an Arrhenius plot (Fig. 7) which contains several 
distinct features. Above ca. 0.3 K the relaxation rate 
is temperature-dependent with $\tau_0 = 3.8 \times 10^{-6}$ s 
and $U_{\rm eff} =$ 10.7 K. 
Below ca. 0.3 K however, the relaxation rate 
is temperature-independent with a relaxation 
rate of $8 \times 10^5$ s indicative of QTM 
between the ground states.  

\begin{figure}
\begin{center}
\includegraphics[width=.4\textwidth]{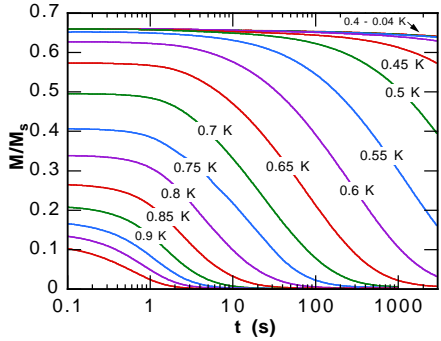}
\caption{
Relaxation measurements at zero applied field at several temperatures.}
\label{relax}
\end{center}
\end{figure}

\begin{figure}
\begin{center}
\includegraphics[width=.4\textwidth]{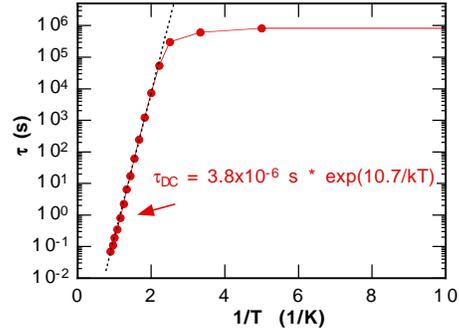}
\caption{
Arrhenius plot of the relaxations times $\tau$ versus 
the inverse temperature.}
\label{tau}
\end{center}
\end{figure}

The above results demonstrate that even weak exchange 
interactions can have a large influence on the quantum 
properties of SMMs. From one viewpoint, each half 
of the [Mn$_4$]$_2$ dimer acts as a field bias on 
its neighbor, shifting the tunnel resonances to new 
positions relative to isolated Mn$_4$ molecules. 
In particular, the latter show a strong QTM step 
at zero field in the hysteresis loop~\cite{Aubin98b}, 
a feature absent for [Mn$_4$]$_2$ (Fig. 2) because 
this would be a double quantum transition with a very 
low probability of occurrence. The absence of tunnelling 
at zero-field is important if SMMs are to be used for 
information storage, and the option is still retained to 
switch on tunnelling, if and when required, by application 
of a field. Thus, future studies will probe this double transition 
at zero field in [Mn$_4$]$_2$, and the corresponding multiple 
quantum transition in higher supramolecular aggregates of SMMs, 
including determining their exact probability of occurrence and 
to what extent this can be controlled by the degree of aggregation, 
variation of exchange coupling strength, and similar modifications. 
From another viewpoint, [Mn$_4$]$_2$ represents an 
unequivocal and unprecedented example of quantum tunnelling 
in a monodisperse antiferromagnet with no uncompensated spin 
(i.e. $S = 0$) in the ground state. Thus, [Mn$_4$]$_2$ and 
related species will prove invaluable for detailed 
studies of this phenomenon.  Finally, the absence of a level 
crossing at zero field also makes the [Mn$_4$]$_2$ dimer a very 
interesting candidate as a qubit for quantum computing, 
because the ground state is the entangled combination of 
the $(9/2,-9/2)$ and $(-9/2,9/2)$ states; the coupling of this $S = 0$ 
system to environmental degrees of freedom should be small, which 
means decoherence effects should also be small. 

In future work, we shall use the Landau-Zener method to 
determine the tunnel splitting in [Mn$_4$]$_2$, and 
apply a transverse field to probe its exact influence on 
QTM rates (we have already confirmed that a transverse 
field increases the tunnelling rate, as expected for QTM). 
The identification of both an antiferromagnetic coupling 
and an exchange-bias effect in [Mn$_4$]$_2$ demonstrates the 
feasibility of employing supramolecular chemistry for 
modulating the quantum physics of SMMs, providing a realistic 
general method to fine-tune the properties of these molecular 
nanoscale materials. This brings closer their use in devices.


\end{document}